\documentclass[journal]{IEEEtran}

\usepackage[utf8]{inputenc}
\usepackage{color}
\usepackage{array}
\usepackage{verbatim}
\usepackage{float}
\usepackage{amsmath}
\usepackage{amsthm}
\usepackage{amssymb}
\usepackage{graphicx}
\usepackage{longtable}
\usepackage{multirow}
\usepackage[unicode=true,
bookmarks=false,
breaklinks=false,pdfborder={0 0 1},colorlinks=false]
{hyperref}
\hypersetup{
	colorlinks,bookmarksopen,bookmarksnumbered,citecolor=blue,urlcolor=blue}
\usepackage{cite}

\floatstyle{ruled}
\newfloat{algorithm}{tbp}{loa}
\providecommand{\algorithmname}{Algorithm}
\floatname{algorithm}{\protect\algorithmname}

\makeatletter
\let\oldforeign@language\foreign@language
\DeclareRobustCommand{\foreign@language}[1]{%
	\lowercase{\oldforeign@language{#1}}}

\let\oldforeign@language\foreign@language
\DeclareRobustCommand{\foreign@language}[1]{%
	\lowercase{\oldforeign@language{#1}}}

\ifCLASSINFOpdf
\else
\fi

\makeatletter

\newtheorem{rem}{Remark}

\newtheorem{assum}{Assumption}

\begin{document}
	\bstctlcite{IEEEexample:BSTcontrol}

\title{Nonlinear Attitude Filter on SO(3): Fast Adaptation and Robustness}

\author{Ajay Singh, Trenton S. Sieb, James H. Howe, and~Hashim~A.~Hashim\\
	Software Engineering\\
	Department of Engineering and Applied Science\\
	Thompson Rivers University,	Kamloops, British Columbia, Canada, V2C-0C8\\
	ludhera17@mytru.ca, siebt19@mytru.ca, howej15@mytru.ca, and hhashim@tru.ca
	\thanks{This work was supported in part by Thompson Rivers University Internal
		research fund, RGS-2020/21 IRF, \# 102315.}
	
}


\maketitle

\begin{abstract}
Nonlinear attitude filters have been recognized to have simpler structure
and better tracking performance when compared with Gaussian attitude
filters and other methods of attitude determination. A key element
of nonlinear attitude filter design is the selection of error criteria.
The conventional design of nonlinear attitude filters has a trade-off
between fast adaptation and robustness. In this work, a new functional
approach based on fuzzy rules for on-line continuous tuning of the
nonlinear attitude filter adaptation gain is proposed. The input and
output membership functions are optimally tuned using artificial bee
colony optimization algorithm taking into account both attitude error
and rate of change of attitude error. The proposed approach results
of high adaptation gain at large error and small adaptation gain at
small error. Thereby, the proposed approach allows fast convergence
properties with high measures of robustness. The simulation results
demonstrate that the proposed approach offers robust and high convergence
capabilities against large error in initialization and uncertain measurements. 
\end{abstract}


\IEEEpeerreviewmaketitle{}

\section{Introduction}

\IEEEPARstart{E}{stimation} of rigid-body orientation in the 3D space is an indispensable
process in robotics and engineering applications such as unmanned
aerial vehicles, mobile robots, underwater vehicles, radar, or satellites
\cite{hashim2018SO3Stochastic,hashim2020SE3Stochastic,hashim2019SO3Det,mahony2008nonlinear,hashim2018Conf1,markley2003attitude,mohamed2019filters,Hashim2020SLAMIEEELetter}.
The orientation of the rigid-body can be reconstructed mathematically
given at least two inertial observations and their body-frame measurements,
for instance, utilizing QUEST computations and singular value decomposition
(SVD) \cite{shuster1981three,markley1988attitude}. In any case, body-frame
estimations are contaminated with unknown constant bias and random
noise elements and the static solutions in \cite{shuster1981three,markley1988attitude}
give unreasonable outcomes, particularly if the moving body is equipped
with cheap measurement units, such as inertial measurements units
(IMUs) \cite{hashim2018SO3Stochastic,hashim2019SO3Det,mahony2008nonlinear,markley2003attitude}.

Over the last two decades, a surprising effort has been done to accomplish
higher estimation capabilities through Gaussian filters. Gaussian
attitude filters includes the Extended Kalman Filter (EKF) in \cite{lefferts1982kalman},
novel Kalman filter in \cite{choukroun2006novel}, Multiplicative
extended Kalman filter \cite{markley2003attitude}, and others. A
more recent survey of Gaussian attitude filters are described in \cite{hashim2018SO3Stochastic}.
Nonlinear deterministic attitude filters have better performance,
and demand less computational energy when contrasted with Gaussian
filters \cite{hashim2018SO3Stochastic,mahony2008nonlinear}. In addition,
they are simpler in derivation. Therefore, nonlinear attitude filters
got considerable attention \cite{hashim2018SO3Stochastic,mahony2008nonlinear,hashim2020SO3Wiley}.

The need for attitude filters that are robust against ambiguity in
estimation sensors, particularly with the advancement in low cost
IMUs, contributed to the improvement of nonlinear attitude filters.
These filters can be effectively fitted knowing a rate gyroscope measurement
and at least two vectorial estimations taken, for example, by IMUs.
The nonlinear attitude filter is accomplished by means of careful
selection of the attitude error function. While the chosen error function
in \cite{mahony2008nonlinear} experienced slight alterations in other
works, overall functioning was essentially unchanged. The main issue
of the error function in \cite{mahony2008nonlinear} is the slow convergence,
particularly with large initial attitude error. Other attitude error
functions were proposed to improve the transient performance as well
as the robustness factor, for instance \cite{hashim2019SO3Det,hashim2020SO3Wiley}.

Fuzzy logic controller (FLC) is classified as an intelligent approach
which showed significant solutions in a wide range of control applications,
for instance, adaptively tuned filter of an $\mathcal{L}_{1}$ adaptive
controller \cite{hashim2015L1} and adaptive fuzzy controller for
mobile robots \cite{shi2019fuzzy}. Evolutionary techniques witnessed
rapid developments over the last few decades and they have the potential
to be an optimal fit for various control applications such as artificial
bee colony (ABC) which was proposed as a global search technique in
\cite{karaboga2008performance}. Also, they have essential role in
data mining \cite{eltoukhy2019data,eltoukhy2019robust,eltoukhy2018joint}.
The necessity to tune originally fixed coefficients of controllers
and filters has been widely used in various applications, such as
\cite{hashim2015L1,yu2017fuzzy}.

To this end, this study proposes fuzzy tuning the gain of the nonlinear
attitude filter, where the fuzzy input and output membership functions
are optimized by ABC taking into account the attitude error and its
rate of change. The FLC-based tuning, is on-line and carried out during
operation. ABC identifies the optimal values of input and output membership
functions through off-line tuning. FLC is introduced to improve the
trade-off between robustness and fast convergence. The gain of the
nonlinear attitude filter is dynamically tuned, allowing for better
performance. In fact, the proposed approach allows the dilemma of
fast adaptation and convergence response to be solved. The method
is simpler and can be easily implemented in comparison with the literature.

The rest of the paper is composed as follows: Section \ref{sec:Problem_Formulation}
gives a short overview of the numerical and mathematical representation,
$SO\left(3\right)$ parameterization, articulates the attitude problem,
demonstrates the estimator structure and error criteria, and presents
the nonlinear structure of the attitude filter. Section \ref{sec:Filter_Strategy}
presents the proposed filter strategy which includes a brief introduction
of the artificial bee colony algorithm, fuzzy logic controller, and
diagram of the implementation process. Section \ref{sec:Results}
presents the obtained results and validates the robustness of the
proposed filters. At long last, Section \ref{sec:SO3PPF_Conclusion}
summarizes the work with finished up comments.

\section{Preliminaries \& Problem Formulation \label{sec:Problem_Formulation}}

The objective of this section is to introduce 1) attitude preliminaries,
2) the attitude estimation problem, 3) the available body-frame measurements,
4) error criteria, and 5) nonlinear filter design.

\subsection{Preliminaries}

In this paper $\left\{ \mathcal{B}\right\} $ denotes body-frame of
a reference and $\left\{ \mathcal{I}\right\} $ denotes inertial-frame
of a reference. $\mathbb{R}_{+}$ is the set of non-negative real
numbers, $\mathbb{R}^{p\times q}$ is real $p\times q$ dimensional
space. $\left\Vert x\right\Vert =\sqrt{x^{\top}x}$ refers to the
Euclidean norm of $x\in\mathbb{R}^{p}$. $\mathbf{I}_{p}$ refers
to a $p$-by-$p$ identity matrix. $R\in\left\{ \mathcal{B}\right\} $
denotes an orientation of a rigid-body in the space which is commonly
know as attitude. Define $SO\left(3\right)$ as the Special Orthogonal
Group. The orientation of a rigid-body in space is defined by
\[
SO\left(3\right)=\left\{ \left.R\in\mathbb{R}^{3\times3}\right|RR^{\top}=R^{\top}R=\mathbf{I}_{3}\text{, }{\rm det}\left(R\right)=+1\right\} 
\]
such that $\mathbf{I}_{3}$ is an identity matrix of dimension $3$
and ${\rm det\left(\cdot\right)}$ is a determinant. The Lie-algebra
associated with $SO\left(3\right)$ is known by $\mathfrak{so}\left(3\right)$
and is represented by
\[
\mathfrak{so}\left(3\right)=\left\{ \left.X\in\mathbb{R}^{3\times3}\right|X^{\top}=-X\right\} 
\]
where $X$ is a skew symmetric matrix. Consider the map $\left[\cdot\right]_{\times}:\mathbb{R}^{3}\rightarrow\mathfrak{so}\left(3\right)$
to be
\[
\left[x\right]_{\times}=\left[\begin{array}{ccc}
0 & -x_{3} & x_{2}\\
x_{3} & 0 & -x_{1}\\
-x_{2} & x_{1} & 0
\end{array}\right]\in\mathfrak{so}\left(3\right),\hspace{1em}x=\left[\begin{array}{c}
x_{1}\\
x_{2}\\
x_{3}
\end{array}\right]
\]
Define $\left[x\right]_{\times}y=x\times y$ where $\times$ is the
cross product $\forall x,y\in\mathbb{R}^{3}$. Let the inverse map
of $\left[\cdot\right]_{\times}$ be defined by $\mathbf{vex}:\mathfrak{so}\left(3\right)\rightarrow\mathbb{R}^{3}$
\begin{equation}
\mathbf{vex}\left(\left[x\right]_{\times}\right)=x\in\mathbb{R}^{3}\label{eq:Attit_VEX}
\end{equation}
The anti-symmetric projection operator on the Lie-algebra $\mathfrak{so}\left(3\right)$
is given by $\boldsymbol{\mathcal{P}}_{a}$ with $\boldsymbol{\mathcal{P}}_{a}:\mathbb{R}^{3\times3}\rightarrow\mathfrak{so}\left(3\right)$
such that
\begin{equation}
\boldsymbol{\mathcal{P}}_{a}\left(K\right)=\frac{1}{2}\left(K-K^{\top}\right)\in\mathfrak{so}\left(3\right),\,K\in\mathbb{R}^{3\times3}\label{eq:Attit_Pa}
\end{equation}
Additionally, the identity below will be used throughout the paper
\begin{align}
{\rm Tr}\left\{ K\left[x\right]_{\times}\right\} = & {\rm Tr}\left\{ \boldsymbol{\mathcal{P}}_{a}\left(K\right)\left[x\right]_{\times}\right\} =-2\mathbf{vex}\left(\boldsymbol{\mathcal{P}}_{a}\left(K\right)\right)^{\top}x\nonumber \\
& \hspace{1em}K\in\mathbb{R}^{3\times3},x\in{\rm \mathbb{R}}^{3}\label{eq:SO3PPF_Identity7}
\end{align}
where ${\rm Tr}\left\{ \cdot\right\} $ denotes trace. For more details, visit \cite{hashim2018SO3Stochastic,hashim2020SO3Wiley}.

\subsection{Attitude Dynamics and Measurements\label{subsec:SE3PPF_Pose-Kinematics}}

\begin{figure}[h]
	\centering{}\includegraphics[scale=0.29]{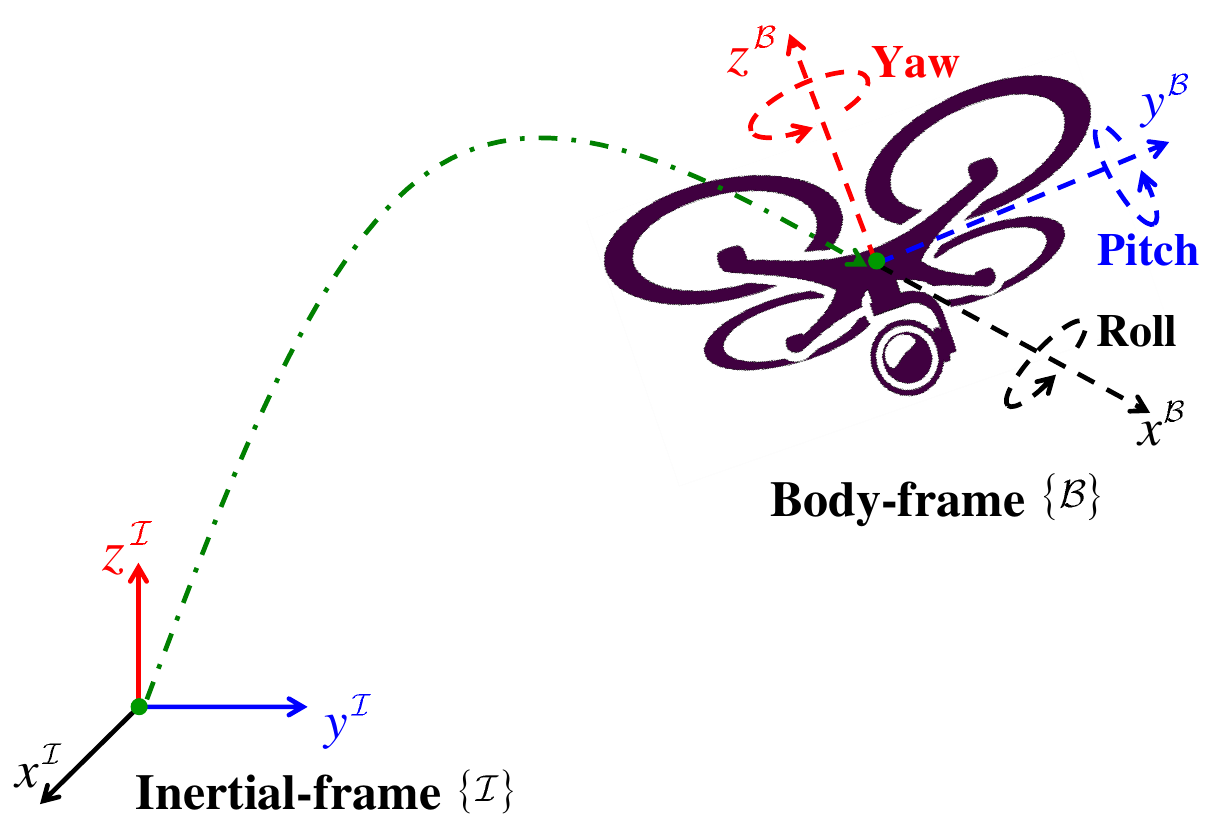}\includegraphics[scale=0.5]{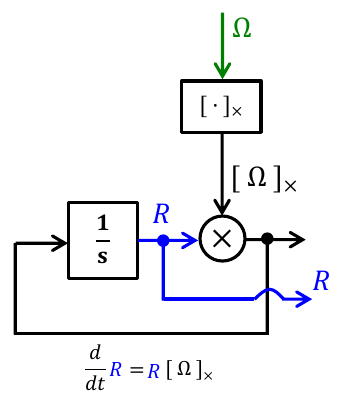}\caption{Attitude estimation problem \cite{hashim2018SO3Stochastic}.}
	\label{fig:SE3PPF_1} 
\end{figure}

The attitude of a rigid-body is described by $R\in SO\left(3\right)$.
Note that $R\in\left\{ \mathcal{B}\right\} $. The attitude estimation
problem of a rigid-body is depicted in Fig. \ref{fig:SE3PPF_1}.

Consider the superscripts $\mathcal{B}$ and $\mathcal{I}$ to be
the components associated with $\left\{ \mathcal{B}\right\} $ and
$\left\{ \mathcal{I}\right\} $, respectively. The orientation can
be represented by $n$ known observations in $\left\{ \mathcal{I}\right\} $
and their measurements in $\left\{ \mathcal{B}\right\} $. The $i$th
vector measurement is 
\begin{equation}
{\rm v}_{i}^{\mathcal{B}}=R^{\top}{\rm v}_{i}^{\mathcal{I}}+b_{i}^{\mathcal{B}}+n_{i}^{\mathcal{B}}\in\mathbb{R}^{3}\label{eq:Attit_Vect_R}
\end{equation}
where ${\rm v}_{i}^{\mathcal{I}}$ is the known $i$th observation,
$b_{i}^{\mathcal{B}}$ is unknown bias, and $n_{i}^{\mathcal{B}}$
is unknown random noise for all $i=1,2,\ldots,n$. The vectors ${\rm v}_{i}^{\mathcal{I}}$
and ${\rm v}_{i}^{\mathcal{B}}$ in Eq. \eqref{eq:Attit_Vect_R} can
be normalized as below:
\begin{equation}
\upsilon_{i}^{\mathcal{I}}=\frac{{\rm v}_{i}^{\mathcal{I}}}{\left\Vert {\rm v}_{i}^{\mathcal{I}}\right\Vert }\in\left\{ \mathcal{I}\right\} ,\hspace{1em}\upsilon_{i}^{\mathcal{B}}=\frac{{\rm v}_{i}^{\mathcal{B}}}{\left\Vert {\rm v}_{i}^{\mathcal{B}}\right\Vert }\in\left\{ \mathcal{B}\right\} \label{eq:Attit_Vector_norm}
\end{equation}

\begin{assum}
	\label{Assum:SE3STCH_1} (Attitude observability) At least two non-collinear
	vectors in Eq. \eqref{eq:Attit_Vector_norm} have to be accessible
	in order to establish the attitude of a rigid-body.
\end{assum}
The orientation dynamics of a rigid-body are defined by
\begin{align}
\dot{R} & =R\left[\Omega\right]_{\times}\label{eq:Attit_R_Dynamics}
\end{align}
where $\Omega$ is the true angular velocity and its measurement is
defined by
\begin{align}
\Omega_{m} & =\Omega+b+n\in\left\{ \mathcal{B}\right\} \label{eq:Attit_Angular}
\end{align}
with $b\in\mathbb{R}^{3}$ being unknown constant bias and $n\in\mathbb{R}^{3}$
being unknown random noise. For more details visit \cite{hashim2018SO3Stochastic,hashim2020SO3Wiley}.
Let $\hat{R}$ denote the estimate of $R$. The aim attitude filters
is to drive $\hat{R}\rightarrow R$ asymptotically as fast as possible.
Accordingly, define the error between body-frame and estimator-frame
as
\begin{equation}
\tilde{R}=R^{\top}\hat{R}\label{eq:Attit_R_error}
\end{equation}
Additionally, define the error in bias estimation as
\begin{align}
\tilde{b} & =b-\hat{b}\label{eq:Attit_b_tilde}
\end{align}

\subsection{Nonlinear Filter Design \label{subsec:Filter_Design}}

Consider the estimate of $\upsilon_{i}^{\mathcal{B}}$ to be
\begin{equation}
\hat{\upsilon}_{i}^{\mathcal{B}}=\hat{R}^{\top}\upsilon_{i}^{\mathcal{I}}\label{eq:SO3_PPF_STCH_vB_hat}
\end{equation}
Also, define
\[
M^{\mathcal{B}}=\sum_{i=1}^{n}s_{i}R^{\top}\upsilon_{i}^{\mathcal{I}}\left(\upsilon_{i}^{\mathcal{I}}\right)^{\top}R
\]
Due to the fact that $\left[\alpha\times\beta\right]_{\times}=\beta\alpha^{\top}-\alpha\beta^{\top},\forall\alpha,\beta\in{\rm \mathbb{R}}^{3}$,
one obtains
\begin{align*}
\left[\sum_{i=1}^{n}\frac{s_{i}}{2}\hat{\upsilon}_{i}^{\mathcal{B}}\times\upsilon_{i}^{\mathcal{B}}\right]_{\times} & =\sum_{i=1}^{n}\frac{s_{i}}{2}\left(\upsilon_{i}^{\mathcal{B}}\left(\hat{\upsilon}_{i}^{\mathcal{B}}\right)^{\top}-\hat{\upsilon}_{i}^{\mathcal{B}}\left(\upsilon_{i}^{\mathcal{B}}\right)^{\top}\right)\\
& =\frac{1}{2}R^{\top}M^{\mathcal{I}}R\tilde{R}-\frac{1}{2}\tilde{R}^{\top}R^{\top}M^{\mathcal{I}}R\\
& =\boldsymbol{\mathcal{P}}_{a}(M^{\mathcal{B}}\tilde{R})
\end{align*}
with the aid of Eq. \eqref{eq:Attit_VEX}, one has
\begin{equation}
\mathbf{vex}\left(\boldsymbol{\mathcal{P}}_{a}(M^{\mathcal{B}}\tilde{R})\right)=\sum_{i=1}^{n}\frac{s_{i}}{2}\hat{\upsilon}_{i}^{\mathcal{B}}\times\upsilon_{i}^{\mathcal{B}}\label{eq:SO3_PPF_STCH_VEX_VM}
\end{equation}
where $s_{i}$ denotes sensor confidence. Also, one can find \cite{hashim2019SO3Det,hashim2020SO3Wiley}
\[
{\rm Tr}\left\{ \tilde{R}\right\} ={\rm Tr}\left\{ \left(\sum_{i=1}^{n}s_{i}\upsilon_{i}^{\mathcal{B}}\left(\upsilon_{i}^{\mathcal{B}}\right)^{\top}\right)^{-1}\sum_{i=1}^{n}s_{i}\upsilon_{i}^{\mathcal{B}}\left(\hat{\upsilon}_{i}^{\mathcal{B}}\right)^{\top}\right\} 
\]
The filter design in this Section follows the structure in \cite{mahony2008nonlinear}
where the contribution is the introducing adaptively tuned again rather
than fixed gain. Consider the following filter design
\begin{equation}
\begin{cases}
\dot{\hat{R}}= & \hat{R}\left[\Omega_{m}-\hat{b}-W\right]_{\times},\quad\hat{R}\left(0\right)=\hat{R}_{0}\\
\dot{\hat{b}}= & \frac{\gamma}{2}{\rm vex}(\boldsymbol{\mathcal{P}}_{a}(M^{\mathcal{B}}\tilde{R}))\\
W= & \boldsymbol{K}{\rm vex}(\boldsymbol{\mathcal{P}}_{a}(M^{\mathcal{B}}\tilde{R}))
\end{cases}\label{eq:Attit_Filter}
\end{equation}
with $\boldsymbol{K}=1+k_{\text{op}}\in\mathbb{R}_{+}$, $k_{\text{op}}$
being a positive constant to be designed in the following Section,
$\gamma$ being a positive constant, and $\hat{b}$ being the estimates
of $b$. One easily obtains \cite{mahony2008nonlinear}
\begin{align*}
\dot{M}^{\mathcal{B}} & =-\left[\Omega\right]_{\times}M^{\mathcal{B}}+M^{\mathcal{B}}\left[\Omega\right]_{\times}
\end{align*}
also
\begin{align*}
\dot{\tilde{R}} & =\tilde{R}\left[\tilde{b}-W\right]_{\times}
\end{align*}
Due to the fact that $\boldsymbol{K}\geq1$ and recalling the identity
in Eq. \eqref{eq:SO3PPF_Identity7}, the proof of stability in \cite{mahony2008nonlinear}
holds for the filter in Eq. \eqref{eq:Attit_Filter}. Recall the identity
in Eq. \eqref{eq:SO3PPF_Identity7} and consider the following Lyapunov
candidate function \cite{mahony2008nonlinear}
\begin{align*}
V= & {\rm Tr}\left(\mathbf{I}_{3}-M^{\mathcal{B}}\tilde{R}\right)+\frac{1}{2\gamma}\tilde{b}^{\top}\tilde{b}
\end{align*}
One has
\begin{align*}
\dot{V}= & -{\rm Tr}\left\{ M^{\mathcal{B}}\dot{\tilde{R}}\right\} -{\rm Tr}\left\{ M^{\mathcal{B}}\tilde{R}\right\} -\frac{1}{\gamma}\tilde{b}^{\top}\dot{\hat{b}}\\
= & \frac{1}{2}{\rm vex}(\boldsymbol{\mathcal{P}}_{a}(M^{\mathcal{B}}\tilde{R}))^{\top}\left(\tilde{b}-W\right)-\frac{1}{\gamma}\tilde{b}^{\top}\dot{\hat{b}}\\
= & -\frac{1}{2}\boldsymbol{K}\left\Vert {\rm vex}(\boldsymbol{\mathcal{P}}_{a}(M^{\mathcal{B}}\tilde{R}))\right\Vert ^{2}
\end{align*}

\begin{rem}
	\label{rem:Remark2}\cite{hashim2019SO3Det,hashim2020SO3Wiley} The
	classic design of nonlinear filters on $SO\left(3\right)$ \cite{mahony2008nonlinear}
	selects the gain $\boldsymbol{K}$ as a positive constant. The remarkable
	weakness of such approach is that smaller values of $\boldsymbol{K}$
	lead to slower transient performance with high measures of robustness
	in the steady-state (less oscillatory performance). On the other hand,
	greater value of $\boldsymbol{K}$ results in faster transient performance
	with less robustness measures in the steady-state (higher oscillation).
\end{rem}
Motivated by the discussion in Remark \ref{rem:Remark2}, the objective
is to find the optimal tuning strategy of the filter gain $\boldsymbol{K}$
which could result in improving 1) the convergence capabilities and
2) robustness.

\section{Proposed Filter Strategy\label{sec:Filter_Strategy}}

According to Remark \ref{rem:Remark2}, $\boldsymbol{K}$ has to be
selected to be large enough at a large error and small enough at a
small error. Therefore, fuzzy logic controller (FLC) will be employed
to control the tuning of $\boldsymbol{K}$ relative to the error in
attitude. FLC consists of 1) fuzzification which include the input
membership function, 2) rule base, and 3) defuzzification which includes
the output membership function. To approach effective design of FLC,
the parameters of input and output membership functions will be set
using the artificial bee colony (ABC) algorithm.

\subsection{Artificial Bee Colony\label{subsec:ABC}}

\begin{figure}
	\centering{}\includegraphics[scale=0.8]{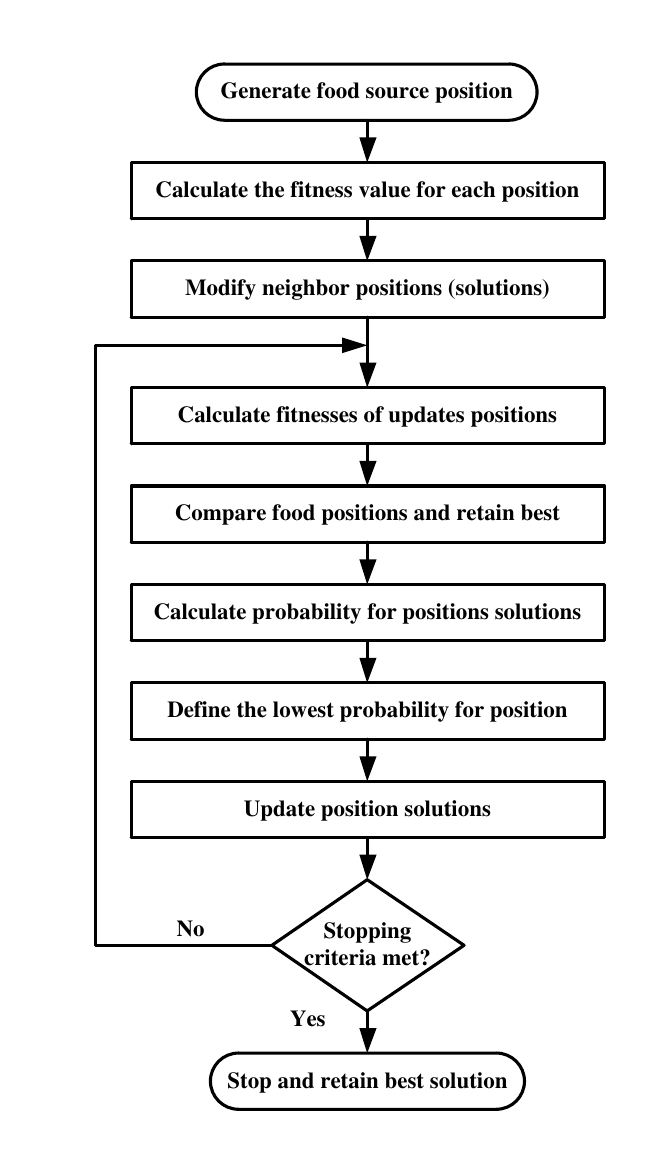}\caption{Graphical illustration of ABC algorithm}
	\label{fig:ABC_algorithm}
\end{figure}

A global optimization approach, intended to imitate the natural behavior
of a colony of bees, called the ABC algorithm was introduced in \cite{karaboga2008performance}.
The algorithm has the colony split into three groups: scout bees are
responsible for searching for new food (solution) sources, employed
bees go to food sources and mark the position of the best one, and
onlooker bees differentiate between good and bad sources brought to
them by the employed bees. Scout bees are tasked with finding new
sources of nectar with no regard to the quality of the food. Their
only responsibility is to facilitate the search process. Employed
bees, however, are searching for the best possible solution in an
area, which they mark the location of, and communicate to the onlookers.
The quantity of nectar is conveyed to the onlookers through a dance
that, based on the duration and speed of shaking, allows them to differentiate
between good and bad sources of food. Together, the employed and onlooker
bees focus on finding the optimum source. The position within the
search space of food sources is modeled as follows:
\begin{equation}
x_{ij}^{new}=x_{ij}^{old}+\alpha\left(x_{ij}^{old}-x_{kj}\right)\label{eq:ABC_Position}
\end{equation}
The quantity of nectar represents the objective function. The probability
that onlooker bees will select a particular food source is modeled
as below:
\begin{equation}
P_{i}=\frac{\mathcal{J}_{i}}{\sum_{i=1}^{N}\mathcal{J}_{i}},\hspace{1em}\forall i=1,2,\ldots,N\label{eq:ABC_Probability}
\end{equation}
The size of the colony is defined by $2N$, $j=1,2,\ldots,P$ where
$P$ is the number of parameters that can be optimized. The related
objective function of $i$th is $\mathcal{J}_{i}$, $k$ is a random
number among the colony size such that $k\in\left\{ 1,2,...,N\right\} $,
and $\alpha$ is a random number where $0\leq\alpha\leq1$. The ABC
algorithm is depicted in Fig. \ref{fig:ABC_algorithm}.

\subsection{Optimal Fuzzy-tuning of Nonlinear Attitude Filter}

FLC uses heuristic and qualitative techniques to enable control of
various applications, hence it is widely used. The development of
FLC, in this work, is to manage nonlinear attitude filter by fine
tuning of the feedback filter gain. This tuning of filter would provide
1) fast convergence of attitude error and 2) improved robustness.

\begin{figure}
	\centering{}\includegraphics[scale=0.7]{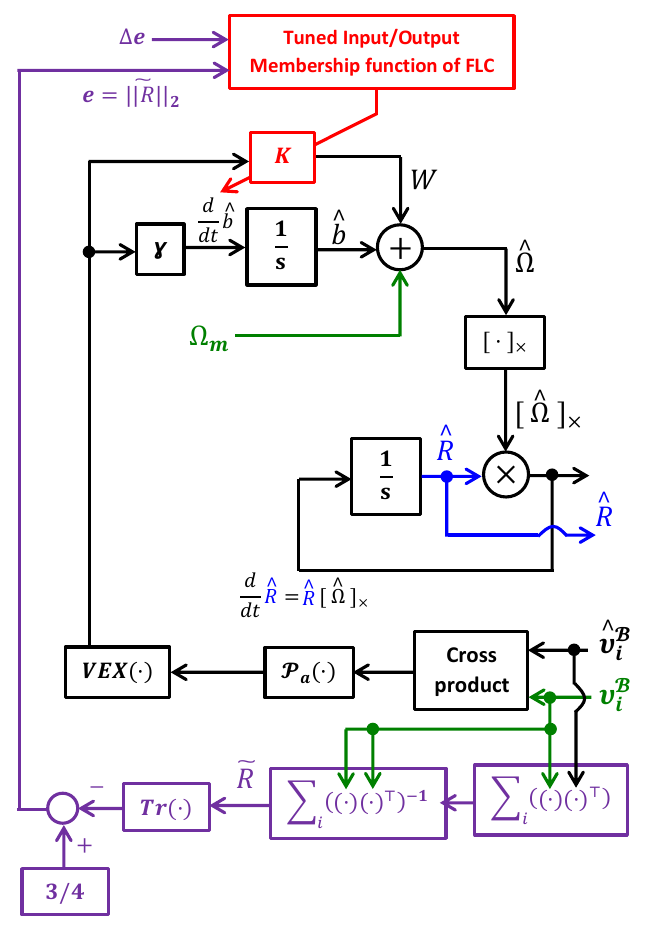}\caption{Graphical illustration of the proposed filter strategy}
	\label{fig:Complete}
\end{figure}

The key objective of this work is the construction of input and output
membership functions for FLC, giving the capability to reduce error.
Through the iterative process, the constraints of the input and output
membership functions were chosen. The triangular membership functions
of fuzzy inputs and output have five linguistic variables. These Linguistic
variables are described as very large ($VL$), large ($L$), medium
($M$), small ($S$) and, very small ($VS$). The optimization of
input and output membership function values are achieved using ABC
in Subsection \ref{subsec:ABC}. See the rule base of the proposed
filter in Table \ref{tab:FLC}. 
\begin{table}
	\caption{\label{tab:FLC}Rule base of FLC.}
	
	\begin{centering}
		\begin{tabular}{|c|c|c|c|c|c|}
			\hline 
			\noalign{\vskip\doublerulesep}
			$\Delta e\backslash e$ & $VL$ & $L$ & $M$ & $S$ & $VS$\tabularnewline[\doublerulesep]
			\hline 
			\hline 
			\noalign{\vskip\doublerulesep}
			$VL$ & $VL$ & $VL$ & $VL$ & $V$ & $V$\tabularnewline[\doublerulesep]
			\hline 
			\noalign{\vskip\doublerulesep}
			$V$ & $VL$ & $VL$ & $VL$ & $V$ & $M$\tabularnewline[\doublerulesep]
			\hline 
			\noalign{\vskip\doublerulesep}
			$M$ & $VL$ & $VL$ & $V$ & $M$ & $M$\tabularnewline[\doublerulesep]
			\hline 
			\noalign{\vskip\doublerulesep}
			$S$ & $VL$ & $VL$ & $M$ & $M$ & $S$\tabularnewline[\doublerulesep]
			\hline 
			\noalign{\vskip\doublerulesep}
			$VS$ & $VL$ & $VL$ & $M$ & $S$ & $VS$\tabularnewline[\doublerulesep]
			\hline 
		\end{tabular}
		\par\end{centering}
\end{table}
The $i$th objective function is selected as below
\begin{equation}
\mathcal{J}_{i}=e_{tr}+e_{ss}=0.3\times\sum_{0\leq t\leq1}e\left(t\right)+\sum_{4\leq t\leq14}e\left(t\right)\label{eq:Cost_function}
\end{equation}
Where $e_{tr}$ refers to the transient time over the period of $0$
to $1$ seconds, while $e_{ss}$ refers to the steady-state error
over the period of $4$ to $14$ seconds for a sampling time of $0.01$
seconds. Also, $0.3$ is a weighting factor. These values were selected
after a set of trials. The input and output membership functions have
constraint values represented by Eq. \eqref{eq:Inp_MMF_Constraints}
and Eq. \eqref{eq:Out_MMF_Constraints}, respectively. Each membership
function is triangular and has three parameters.

\begin{equation}
\begin{cases}
\left[0,0,0\right] & \leq\left[0,0,k_{1}\right]\leq\left[0,0,0.15\right]\\
\left[0,0,0.1\right] & \leq\left[k_{2},k_{3},k_{4}\right]\leq\left[0.2,0.2,0.2\right]\\
\left[0.05,0.1,0.1\right] & \leq\left[k_{5},k_{6},k_{7}\right]\leq\left[0.2,0.25,0.35\right]\\
\left[0.1,0.2,0.2\right] & \leq\left[k_{8},k_{9},k_{10}\right]\leq\left[0.35,0.5,0.7\right]\\
\left[0.2,1,1\right] & \leq\left[k_{11},1,1\right]\leq\left[0.7,1,1\right]
\end{cases}\label{eq:Inp_MMF_Constraints}
\end{equation}
\begin{equation}
\begin{cases}
\left[0,0,0\right] & \leq\left[0,0,k_{12}\right]\leq\left[0,0,15\right]\\
\left[0,5,10\right] & \leq\left[k_{13},k_{14},k_{15}\right]\leq\left[20,20,30\right]\\
\left[5,20,20\right] & \leq\left[k_{16},k_{17},k_{18}\right]\leq\left[20,50,50\right]\\
\left[20,20,40\right] & \leq\left[k_{19},k_{20},k_{21}\right]\leq\left[50,70,90\right]\\
\left[40,100,100\right] & \leq\left[k_{22},100,100\right]\leq\left[70,100,100\right]
\end{cases}\label{eq:Out_MMF_Constraints}
\end{equation}
where $k_{1}$ to $k_{22}$ are parameters of the membership functions
to be optimized using the ABC algorithm with respect to the objective
function in Eq. \eqref{eq:Cost_function} as well as the constraints
in Eq. \eqref{eq:Inp_MMF_Constraints} and Eq. \eqref{eq:Out_MMF_Constraints},
respectively. Fig. \ref{fig:Complete} illustrates the complete diagram
of the proposed filter strategy.

\section{Results and Discussion\label{sec:Results}}

\subsection{Attitude Measurements and Initialization}

Consider the following set of measurements 
\[
\begin{cases}
\Omega_{m} & =\Omega+b+n\text{ (rad/sec)}\\
\Omega & =\left[\sin\left(0.4t\right),\sin\left(0.7t+\frac{\pi}{4}\right),0.4\sin\left(0.3t+\frac{\pi}{2}\right)\right]^{\top}\\
b & =0.1\left[-1,1,0.5\right]^{\top},\hspace{1em}\hspace{1em}n=\mathcal{N}\left(0,0.2\right)\\
{\rm v}_{i}^{\mathcal{B}} & =R^{\top}{\rm v}_{i}^{\mathcal{I}}+b_{i}^{\mathcal{B}}+n_{i}^{\mathcal{B}}\\
{\rm v}_{1}^{\mathcal{I}} & =\frac{1}{\sqrt{3}}\left[1,-1,1\right]^{\top}\\
b_{1}^{\mathcal{B}} & =0.1\left[1,-1,1\right]^{\top},\hspace{1em}\hspace{1em}n_{1}^{\mathcal{B}}=\mathcal{N}\left(0,0.05\right)\\
{\rm v}_{2}^{\mathcal{I}} & =\left[0,0,1\right]^{\top}\\
b_{2}^{\mathcal{B}} & =0.1\left[0,0,1\right]^{\top},\hspace{1em}\hspace{1em}n_{2}^{\mathcal{B}}=\mathcal{N}\left(0,0.05\right)\\
{\rm v}_{3}^{\mathcal{B}} & ={\rm v}_{1}^{\mathcal{B}}\times{\rm v}_{2}^{\mathcal{B}}
\end{cases}
\]
where $n=\mathcal{N}\left(0,0.2\right)$ is a random noise vector
with $0$ mean and standard deviation of $0.2$. Also, for very large
error, consider the following initialization
\[
R\left(0\right)=\mathbf{I}_{3},\hspace{1em}\hat{R}\left(0\right)=\left[\begin{array}{ccc}
-0.0074 & 0.8557 & 0.5175\\
0.8802 & -0.2399 & 0.4094\\
0.4745 & 0.4586 & -0.7514
\end{array}\right]
\]

\subsection{ABC Implementation}

\begin{figure}
	\centering{}\includegraphics[scale=0.24]{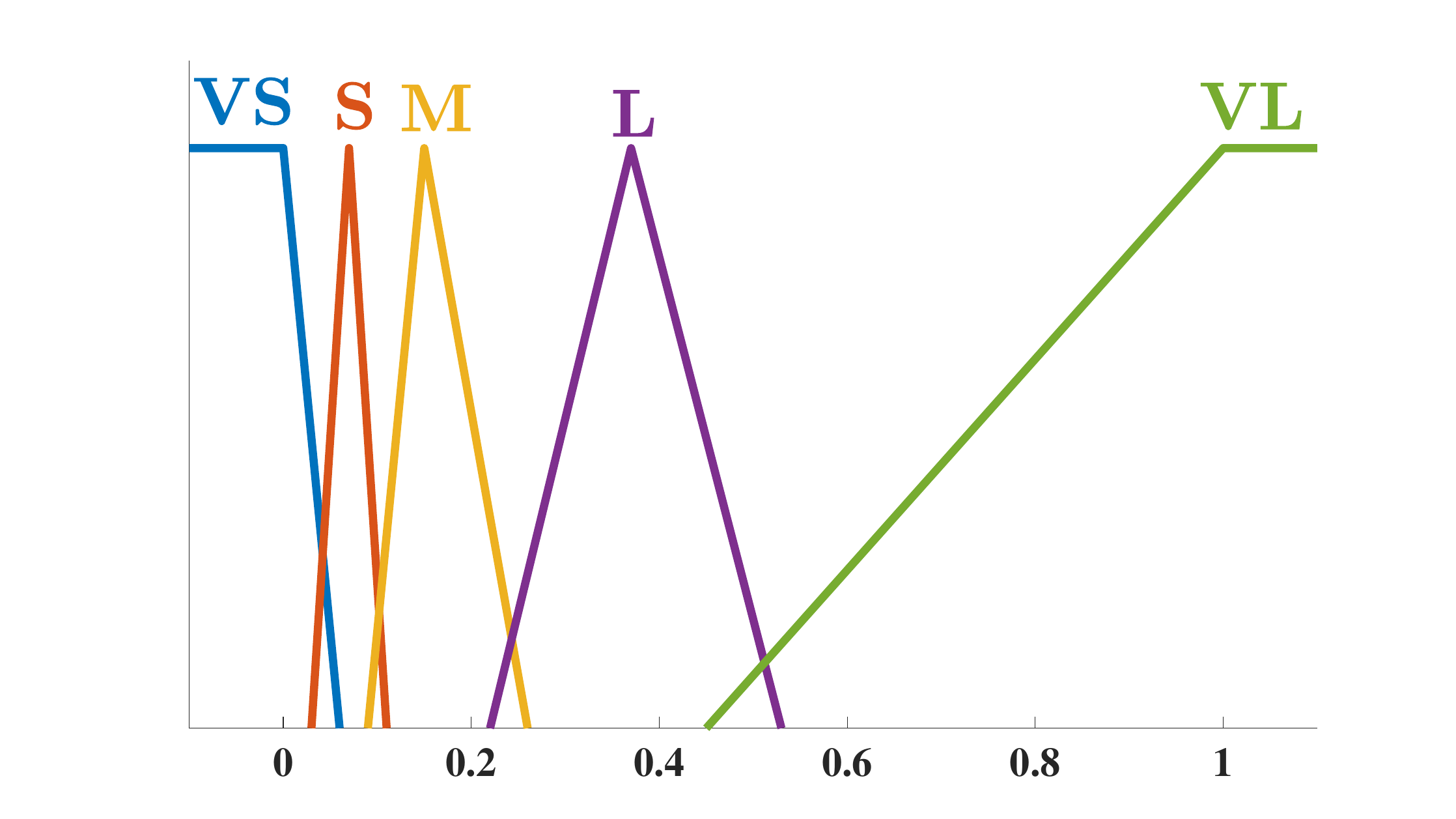}\caption{Error and rate of error membership functions}
	\label{fig:FLC_Inp_MMF}
\end{figure}

\begin{figure}
	\centering{}\includegraphics[scale=0.24]{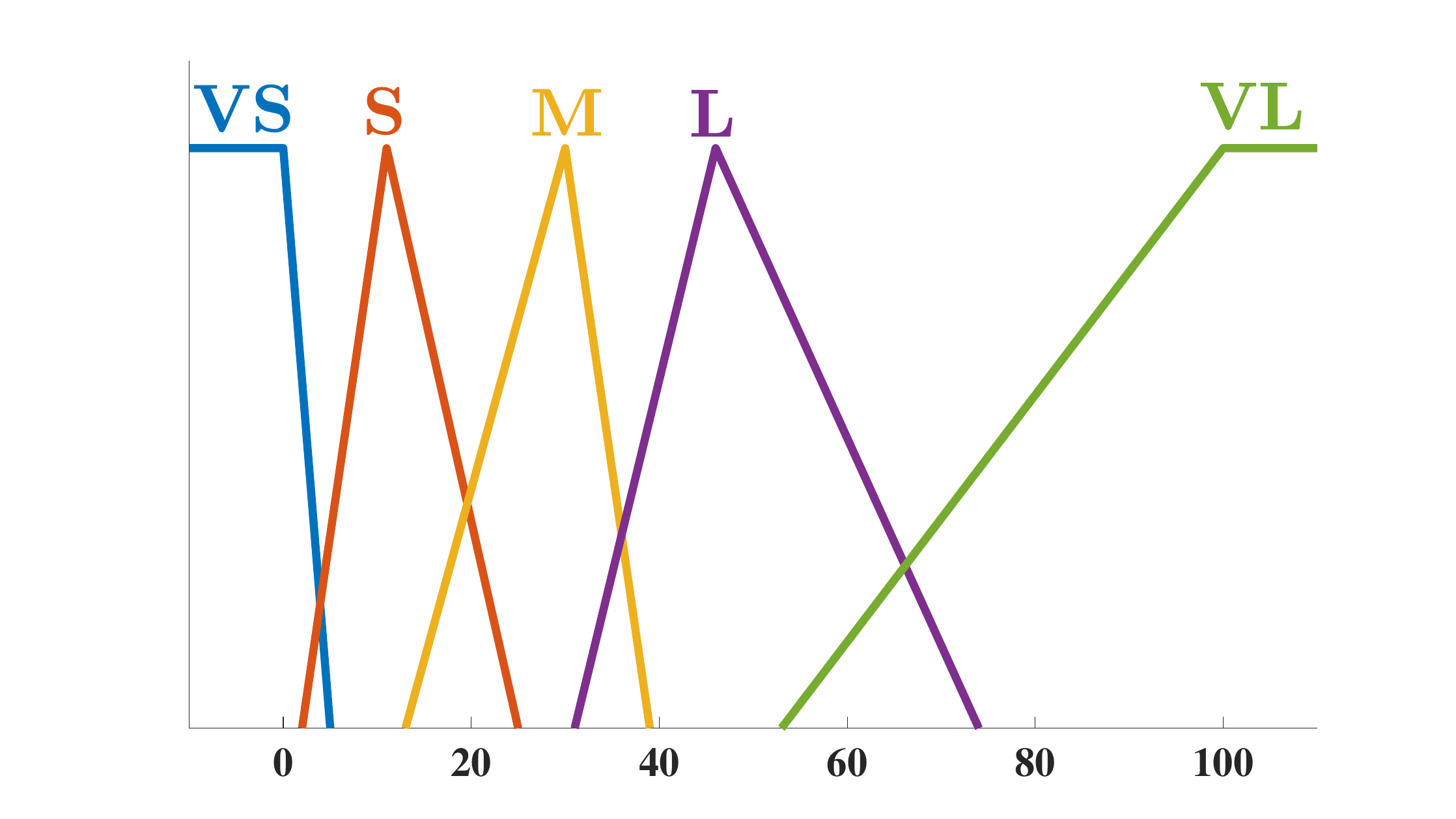}\caption{Output membership function}
	\label{fig:FLC_Out_MMF}
\end{figure}

For implementation, Eq. \eqref{eq:ABC_Position} represents the position
of the food source and Eq. \eqref{eq:ABC_Probability} represents
quality of the food. $N$ is the number of sources to be visited.
According to the ABC algorithm, $N$ represents half of the colony
size. The total number of iterations is 300. The number of sources
to be visited in each iteration are $N=100$. In every source visit,
there are $22$ parameters to be optimized $k_{1}$ to $k_{22}$ given
in Eq. \eqref{eq:Inp_MMF_Constraints} and \eqref{eq:Out_MMF_Constraints}.
Fig. \ref{fig:FLC_Inp_MMF} and \ref{fig:FLC_Out_MMF} illustrate
the optimized input and output membership function after completing
the search process.

\subsection{Results of the Proposed Filter Strategy}

\begin{figure}
	\centering{}\includegraphics[scale=0.36]{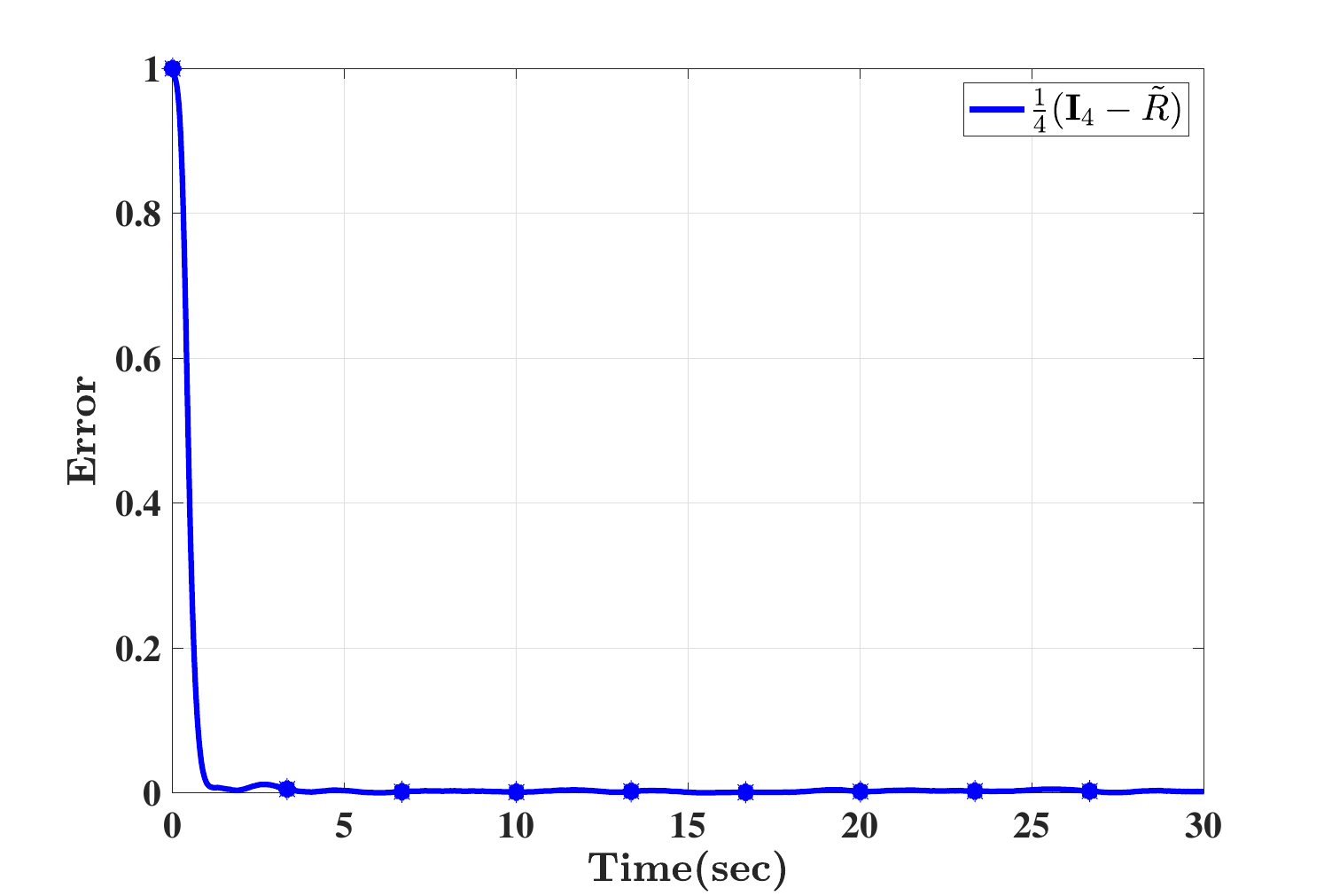}\caption{Normalized Euclidean distance}
	\label{fig:RI2}
\end{figure}

\begin{figure}
	\centering{}\includegraphics[scale=0.28]{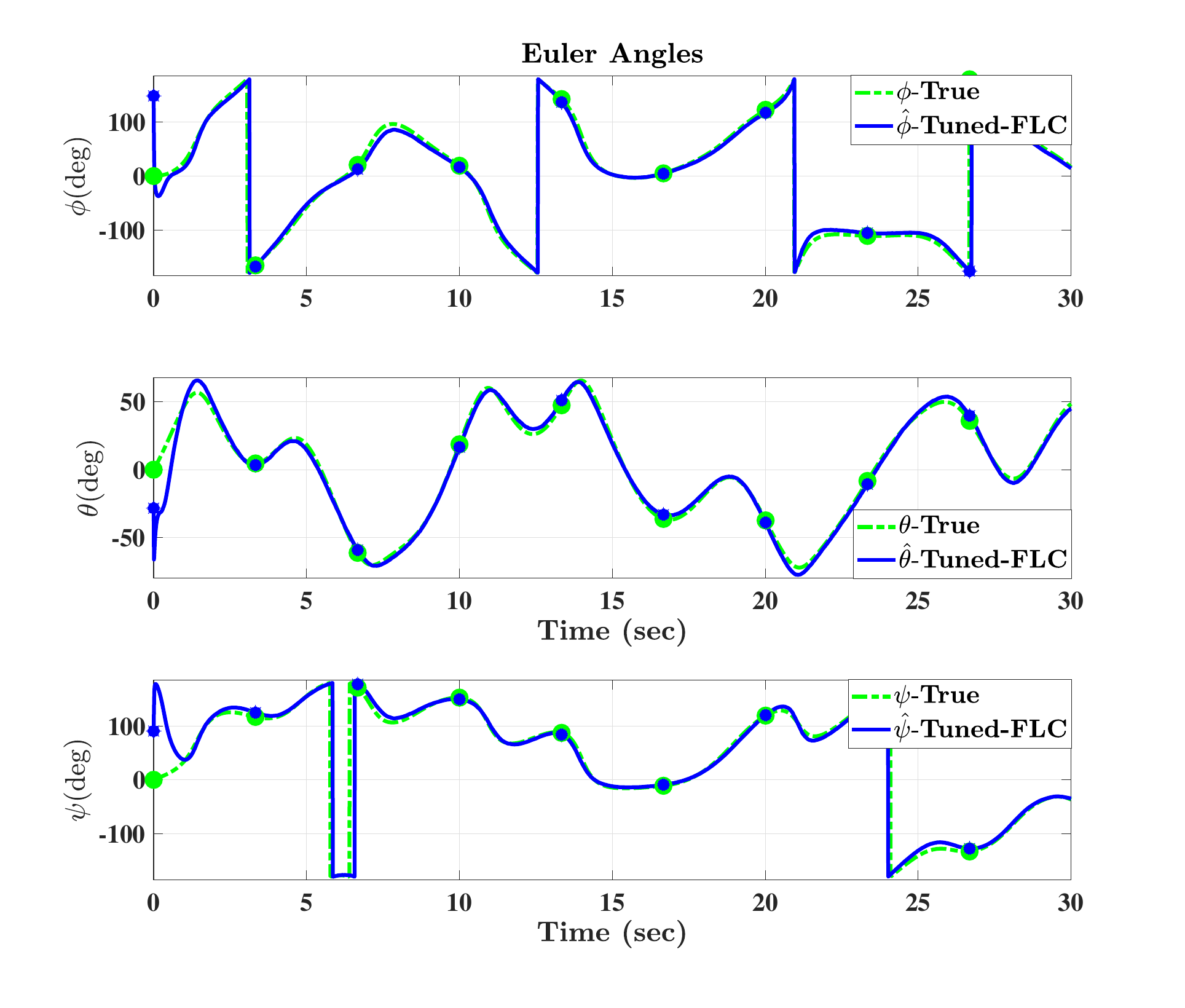}\caption{Euler angles: True vs Estimate (Proposed)}
	\label{fig:Euler}
\end{figure}

Fig. \ref{fig:RI2} depicts smooth and fast convergence of the normalized
Euclidean distance error $||\tilde{R}||_{I}=\frac{1}{4}{\rm Tr}\left\{ \mathbf{I}_{3}-R^{\top}\hat{R}\right\} $.
It is obvious that $||\tilde{R}||_{I}$ started near the unstable
equilibria and settled very close to the origin. The superiority of
the proposed filter can be confirmed by Fig. \ref{fig:Euler}. The
true Euler angles $\left(\phi,\theta,\psi\right)$ are plotted in
Fig. \ref{fig:Euler} versus the estimated angles. Again, Fig. \ref{fig:Euler}
illustrates robust performance of the proposed filter with superior
convergence capabilities.

\section{Conclusion\label{sec:SO3PPF_Conclusion}}

This paper presents a new fuzzy logic controller (FLC) design for
the adaptation gain of the nonlinear attitude filter. The artificial
bee colony (ABC) optimization algorithm has been employed to determine
the optimal variables of the input and output membership functions
of the FLC. The proposed approach tunes the adaptation gain on-line,
which in turn leads to fast adaptation. Additionally, because of the
smooth tuning, the proposed filter maintains a high measure of robustness.
Numerical results reveal fast convergence of the attitude error and
robustness of the proposed filter for the case of large initialization
error and uncertain measurements. There are several directions for
future work. The first direction is to implement the proposed approach
on a real module and compare it against existing techniques in literature.
The second direction is implementing recursive ABC for on-line implementation.
To our knowledge, recursive ABC has not yet been explored within the
area of attitude filters. The third direction is comparing ABC against
other methods of evolutionary techniques.

\section*{Acknowledgment}

The authors would like to thank \textbf{Maria Shaposhnikova} for proofreading
the article.

\bibliographystyle{IEEEtran}
\bibliography{bib_AttitudeFile}

\begin{thebibliography}{10}
\providecommand{\url}[1]{#1}
\csname url@samestyle\endcsname
\providecommand{\newblock}{\relax}
\providecommand{\bibinfo}[2]{#2}
\providecommand{\BIBentrySTDinterwordspacing}{\spaceskip=0pt\relax}
\providecommand{\BIBentryALTinterwordstretchfactor}{4}
\providecommand{\BIBentryALTinterwordspacing}{\spaceskip=\fontdimen2\font plus
\BIBentryALTinterwordstretchfactor\fontdimen3\font minus
  \fontdimen4\font\relax}
\providecommand{\BIBforeignlanguage}[2]{{%
\expandafter\ifx\csname l@#1\endcsname\relax
\typeout{** WARNING: IEEEtran.bst: No hyphenation pattern has been}%
\typeout{** loaded for the language `#1'. Using the pattern for}%
\typeout{** the default language instead.}%
\else
\language=\csname l@#1\endcsname
\fi
#2}}
\providecommand{\BIBdecl}{\relax}
\BIBdecl

\bibitem{hashim2018SO3Stochastic}
H.~A. Hashim, L.~J. Brown, and K.~McIsaac, ``Nonlinear stochastic attitude
  filters on the special orthogonal group 3: Ito and stratonovich,'' \emph{IEEE
  Transactions on Systems, Man, and Cybernetics: Systems}, vol.~49, no.~9, pp.
  1853--1865, 2019.

\bibitem{hashim2020SE3Stochastic}
H.~A. Hashim and F.~L. Lewis, ``Nonlinear stochastic estimators on the special
  euclidean group {SE}(3) using uncertain imu and vision measurements,''
  \emph{IEEE Transactions on Systems, Man, and Cybernetics: Systems}, vol.~PP,
  no.~PP, pp. 1--14, 2020.

\bibitem{hashim2019SO3Det}
H.~A. Hashim, L.~J. Brown, and K.~McIsaac, ``Guaranteed performance of
  nonlinear attitude filters on the special orthogonal group {SO}(3),''
  \emph{IEEE Access}, vol.~7, no.~1, pp. 3731--3745, 2019.

\bibitem{mahony2008nonlinear}
R.~Mahony, T.~Hamel, and J.-M. Pflimlin, ``Nonlinear complementary filters on
  the special orthogonal group,'' \emph{IEEE Transactions on Automatic
  Control}, vol.~53, no.~5, pp. 1203--1218, 2008.

\bibitem{hashim2018Conf1}
H.~A. Hashim, L.~J. Brown, and K.~McIsaac, ``Nonlinear explicit stochastic
  attitude filter on {SO}(3),'' in \emph{Proceedings of the 57th {IEEE}
  conference on {D}ecision and {C}ontrol ({CDC})}, 2018, pp. 1210 --1216.

\bibitem{markley2003attitude}
F.~L. Markley, ``Attitude error representations for kalman filtering,''
  \emph{Journal of guidance, control, and dynamics}, vol.~26, no.~2, pp.
  311--317, 2003.

\bibitem{mohamed2019filters}
H.~A.~H. Mohamed, ``Nonlinear attitude and pose filters with superior
  convergence properties,'' \emph{Ph. D, Western University}, 2019.

\bibitem{Hashim2020SLAMIEEELetter}
H.~A. Hashim, ``Guaranteed performance nonlinear observer for simultaneous
  localization and mapping,'' \emph{IEEE Control Systems Letters}, vol.~5,
  no.~1, pp. 91--96, 2021.

\bibitem{shuster1981three}
M.~D. Shuster and S.~D. Oh, ``Three-axis attitude determination from vector
  observations,'' \emph{Journal of Guidance, Control, and Dynamics}, vol.~4,
  pp. 70--77, 1981.

\bibitem{markley1988attitude}
F.~L. Markley, ``Attitude determination using vector observations and the
  singular value decomposition,'' \emph{Journal of the Astronautical Sciences},
  vol.~36, no.~3, pp. 245--258, 1988.

\bibitem{lefferts1982kalman}
E.~J. Lefferts, F.~L. Markley, and M.~D. Shuster, ``Kalman filtering for
  spacecraft attitude estimation,'' \emph{Journal of Guidance, Control, and
  Dynamics}, vol.~5, no.~5, pp. 417--429, 1982.

\bibitem{choukroun2006novel}
D.~Choukroun, I.~Y. Bar-Itzhack, and Y.~Oshman, ``Novel quaternion kalman
  filter,'' \emph{IEEE Transactions on Aerospace and Electronic Systems},
  vol.~42, no.~1, pp. 174--190, 2006.

\bibitem{hashim2020SO3Wiley}
H.~A. Hashim, ``Systematic convergence of nonlinear stochastic estimators on
  the special orthogonal group {SO}(3),'' \emph{International Journal of Robust
  and Nonlinear Control}, vol.~30, no.~10, pp. 3848--3870, 2020.

\bibitem{hashim2015L1}
H.~A. Hashim, S.~El-Ferik, and M.~A. Abido, ``A fuzzy logic feedback filter
  design tuned with pso for {L}1 adaptive controller,'' \emph{Expert Systems
  with Applications}, vol.~42, no.~23, pp. 9077--9085, 2015.

\bibitem{shi2019fuzzy}
H.~Shi, M.~Xu, and K.-S. Hwang, ``A fuzzy adaptive approach to decoupled visual
  servoing for a wheeled mobile robot,'' \emph{IEEE Transactions on Fuzzy
  Systems}, 2019.

\bibitem{karaboga2008performance}
D.~Karaboga and B.~Basturk, ``On the performance of artificial bee colony (abc)
  algorithm,'' \emph{Applied soft computing}, vol.~8, no.~1, pp. 687--697,
  2008.

\bibitem{eltoukhy2019data}
A.~E. Eltoukhy, Z.~Wang, F.~T. Chan, and X.~Fu, ``Data analytics in managing
  aircraft routing and maintenance staffing with price competition by a
  stackelberg-nash game model,'' \emph{Transportation Research Part E:
  Logistics and Transportation Review}, vol. 122, pp. 143--168, 2019.

\bibitem{eltoukhy2019robust}
A.~E. Eltoukhy, Z.~Wang, F.~T. Chan, S.~Chung, H.-L. Ma, and X.~Wang, ``Robust
  aircraft maintenance routing problem using a turn-around time reduction
  approach,'' \emph{IEEE Transactions on Systems, Man, and Cybernetics:
  Systems}, 2019.

\bibitem{eltoukhy2018joint}
A.~E. Eltoukhy, Z.~Wang, F.~T. Chan, and S.~H. Chung, ``Joint optimization
  using a leader--follower stackelberg game for coordinated configuration of
  stochastic operational aircraft maintenance routing and maintenance
  staffing,'' \emph{Computers \& Industrial Engineering}, vol. 125, pp. 46--68,
  2018.

\bibitem{yu2017fuzzy}
Y.~Yu, Z.~Yang, C.~Han, and H.~Liu, ``Fuzzy adaptive back-stepping sliding mode
  controller for high-precision deflection control of the magnetically
  suspended momentum wheel,'' \emph{IEEE Transactions on Industrial
  Electronics}, vol.~65, no.~4, pp. 3530--3538, 2017.

\end{thebibliography}
\newpage

\end{document}